\documentclass[preprint,showpacs,preprintnumbers,amsmath,amssymb,endfloats*]{revtex4}
\usepackage{graphicx}

\begin{document}
\def \ee {\varepsilon}
\thispagestyle{empty}
{\bf
Comment on ``Application of the Lifshitz theory to poor conductors''
}

The Letter \cite{1} is devoted to the generalization 
of the Lifshitz theory  to spatially nonlocal 
conducting materials. Spatial dispersion is taken into account approximately 
by means of the dielectric permittivities calculated in the random phase
approximation.
Letter \cite{1} recognizes that these permittivities are rigorously
defined only for an infinite medium, but claims that they can be used
for two semispaces separated by a gap as well. Additionally, the specular
reflection of charge carriers on the boundary planes is assumed.
It has been proven, however, that for spatially dispersive materials the
scattering of carriers is neither specular nor diffuse \cite{3}. Thus,
the developed approach is a crude approximation. It does not contain
self-consistent checks  of its accuracy and
could be justified only because of agreement with  experimental data
and fundamental physical principles. Below we show that the claims
of \cite{1} regarding consistency with the experimental data and Nernst's
theorem are based on irregular comparison with the data and 
misinterpretation of relevant physical quantities.

According to \cite{1}, the experimental data of \cite{4}
for the difference Casimir force between an Au sphere and a Si plate
in the presence and in the absence of laser light are equally consistent
with the nonlocal approach and 
``local theory with zero conductivity of Si''.
(Note that in \cite{4} Si conductivity was disregarded only in the absence 
of light.) To prove this, in Fig.~1(a) of \cite{1} the experimental data
[taken from Fig.~1(a) of \cite{5} without a reference] are shown at a 70\%
confidence level. In the same figure the theoretical band for the nonlocal 
approach is obtained from the uncertainty in  $n$, 
$\Delta n=0.4\times 10^{19}\,\mbox{cm}^{-3}$, determined in \cite{4}
at a 95\% confidence level. 
The reader of \cite{1} is not informed about the 
confidence levels used.
Such a mismatch comparison of experiment with theory
is irregular. In Fig.~1(a) of this Comment
the same data are compared with the predictions
of the nonlocal approach and the local theory (the bands between the dashed 
and solid lines, respectively), where the band widths are
determined
 at the same 70\% confidence level, as the errors of the data. 
It is clearly
seen that the local theory is consistent with data, whereas the nonlocal
approach is excluded by the data at a 70\% confidence level. 
[Fig.~1(b)
in \cite{1} shows
both the data and the theoretical bands  at
a 95\% confidence level. However, as stated in \cite{5},  
these  data 
cannot be conclusively compared with the nonlocal approach at such a
high confidence.] 

According to \cite{1}, the nonlocal approach is also applicable to 
metallic test bodies. In Fig.~1(b) we plot as crosses the mean
measured Casimir pressures between two Au plates \cite{6}
at a 95\% confidence compared with computations
using the nonlocal approach (the grey band) and the generalized plasma-like
permittivity (the black band). The widths of both bands are determined at 
a 95\% confidence. 
It can be seen that the nonlocal approach is excluded by the 
data at a 95\% confidence, whereas the local theory is consistent with
them.

According to \cite{1}, the nonlocal approach satisfies the Nernst
theorem, specifically, for ionic conductors possessing an 
activation type conductivity. 
To prove this, the Letter \cite{1} arbitrarily
 separates the
thermal dependence $\sim \exp(-E_a/k_BT)$, where $E_a$ is the activation
energy, from the mobility $\mu$ and attributes it to the ``effective
density of charges, which are able to move".  This transfer of the
temperature dependence from $\mu$ to $n$ is incorrect because 
the commonly used density of charge carriers $n$ producing the effect
of screening in
ionic conductors is an independently measured quantity \cite{7},
which does not vanish with $T$.
In reality, the screening length depends on the standard charge carrier
density $n$ and this results in the violation of Nernst's theorem in the
nonlocal approach to ionic conductors \cite{5}. %\hfill\\[1mm]

\noindent
G.~L.~Klimchitskaya,${}^1$ U.~Mohideen,${}^2$
and V.~M.~Mostepanenko${}^1$ \hfill \\
${}^1$Institute for Theoretical
Physics, Leipzig University,
D-04009, Leipzig, Germany \hfill \\
${}^2$Department of Physics and Astronomy, University of California,
Riverside, CA 92521, USA\hfill \\
PACS numbers: 42.50.Ct, 12.20.Ds, 42.50.Lc, 78.20.Ci

%%%%%%%%%%%%%%
\begin{figure}
\vspace*{-4cm}
\centerline{
\includegraphics{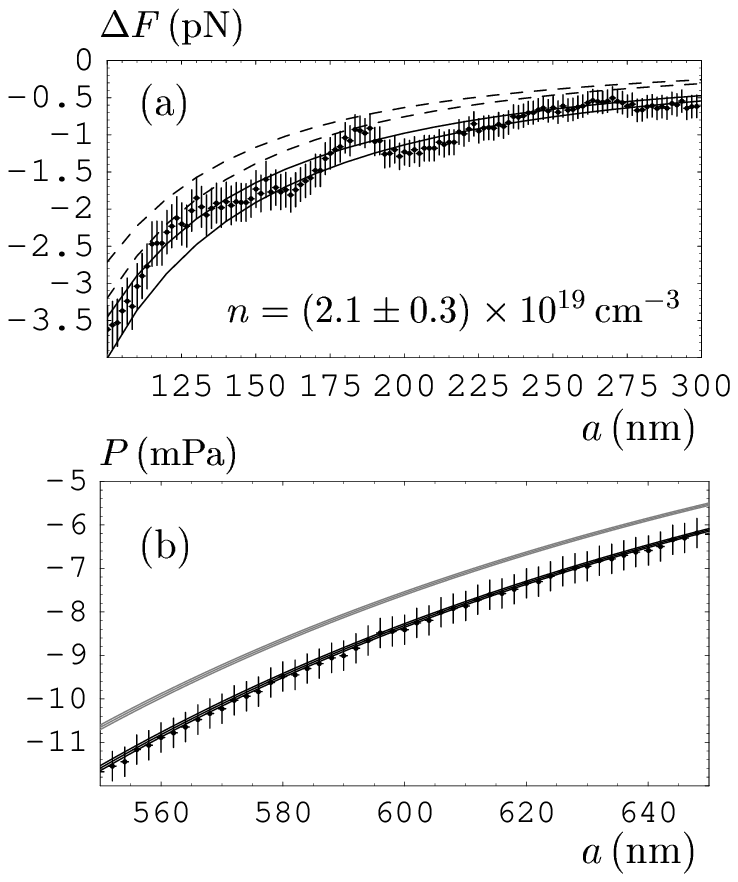}
}
\vspace*{-12cm}
\caption{The measured (a) force differences and (b) pressures are shown 
as crosses versus separation. 
The theoretical bands in (a) lie between the solid and dashed lines,
respectively, and in (b) are indicated as black  and grey stripes. See text for
further discussion.
}
\end{figure}
%%%%%%%%%%%%%%
\end{document}